# PhaseNet: A Deep-Neural-Network-Based Seismic Arrival Time Picking Method

Weiqiang Zhu⋆ and Gregory C. Beroza

*Department of Geophysics, Stanford University*

## SUMMARY

As the number of seismic sensors grows, it is becoming increasingly difficult for analysts to pick seismic phases manually and comprehensively, yet such efforts are fundamental to earthquake monitoring. Despite years of improvements in automatic phase picking, it is difficult to match the performance of experienced analysts. A more subtle issue is that different seismic analysts may pick phases differently, which can introduce bias into earthquake locations. We present a deep-neural-network-based arrival-time picking method called "PhaseNet" that picks the arrival times of both P and S waves. Deep neural networks have recently made rapid progress in feature learning, and with sufficient training, have achieved super-human performance in many applications. PhaseNet uses three-component seismic waveforms as input and generates probability distributions of P arrivals, S arrivals, and noise as output. We engineer PhaseNet such that peaks in probability provide accurate arrival times for both P and S waves, and have the potential to increase the number of S-wave observations dramatically over what is currently available. This will enable both improved locations and improved shear wave velocity models. PhaseNet is trained on the prodigious available data set provided by analyst-labeled P and S arrival times from the Northern California Earthquake Data Center. The dataset we use con-tains more than seven million waveform samples extracted from over thirty years of earthquake recordings. We demonstrate that PhaseNet achieves much higher picking accuracy and recall rate than existing methods.

## 1 INTRODUCTION

Earthquake detection and location are fundamental to seismology. The quality of earthquake catalogs depends critically on both the number and the accuracy of arrival time measurements. Earthquake arrival time measurement, or phase picking, is often carried out by network analysts who base their phase pick on expert judgment and years of experience. As the rate of seismometer deployment continues to accelerate; however, it is becoming increasingly difficult to keep up with the data flow. This is particularly true for dense networks in areas of particular interest or concern that now may contain 1000s of sensors. Phase pickers are particularly challenged by S waves, because they are not the first arriving waves, and they emerge from the scattered waves of the P coda. S wave arrival times are particularly useful because they can be used to reduce the depth-origin trade-off that can afflict earthquake locations based on P waves alone, and because S-wave structure is important for strong ground motion prediction.

Decades of research has been devoted to automatic phase picking, including: methods based on amplitude, standard deviation or energy; statistical methods and shallow neural networks. The short-term average/long-term average (STA/LTA) method (Allen 1978) is commonly used and tracks the ratio of energy in a short-term window with that in a long-term window. Peaks above a threshold mark impulsive P or S wave arrivals. This method is efficient, often effective, but susceptible to noise and has low accuracy for arrival times, particularly for shear waves. Baer & Kradolfer (1987) improved the STA/LTA method using the envelope as characteristic function. Sleeman & Van Eck (1999) applied joint autoregressive (AR) modeling of the noise and seismic signal and used the Akaike Information Criterion (AIC) to determine the onset of a seismic signal. Approaches based on higher-order statistics (HOS), including kurtosis and skewness, were developed to identify the transition from Gaussianity to non-Gaussianity, which coincides with the onset of the seismic event, even in the presence of noise (Saragiotis et al. 2002; Küperkoch et al. 2010). Traditional shallow neural networks were tested by Gentili & Michelini (2006), based on four manually defined features: variance, absolute value of skewness, kurtosis and a combination of skewness and kurtosis predicted based on sliding windows. While most phase picking algorithms focus on P waves, Ross & Ben-Zion (2014) utilized polarization analysis to distinguish between P and S waves primarily to improve S-wave arrival time measurements. Despite the substantial efforts outlined above, the accuracy of automated phase picking algorithms lags that of experienced analysts. This is attributable to the fact that earthquake waveforms are highly complex due to multiple effects including: source mechanism, stress drop, scattering, site-effects, phase conversions, and interference from a multitude of noise sources. Traditional automated methods use manually defined features that require careful data processing, like band-pass filtering and setting an activation threshold.

⋆ zhuwq@stanford.edu

In this paper, we present a deep neural network algorithm, PhaseNet, for seismic phase picking. Instead of using manually defined features, deep neural networks learn the features from labeled data, both noise and signal, which proves a powerful advantage for complex seismic waveforms. The network is trained on the substantial catalog of available P and S arrival-times picked by experienced analysts. Unfiltered three-component seismic waveforms are the input to PhaseNet, which is trained to output three probability distributions: P wave, S wave, and noise. The neural network is trained on the target probability distributions of known earthquake waveforms. Peaks in the P wave and S wave probability distributions are designed to correspond to the predicted P and S arrival times. We demonstrate that PhaseNet provides high accuracy and recall rate for both P and S picks, and achieves significant improvement compared with a traditional STA/LTA method. PhaseNet has the potential to provide comprehensive, superior performance for standard earthquake monitoring.

## 2 DATA

Seismological archives include tremendous numbers of manually picked P and S wave arrivals, which represent an exceptionally rich training set of labeled data that is ideal for deep learning (Figure 1). In this paper, we gathered available digital seismic waveform data based on the Northern California Earthquake Data Center Catalog (NCEDC 2014). We use three-component data that have both P and S arrival times. This leaves us 779,514 recordings in the dataset. We use stratified sampling based on stations to divide this dataset into training, validation and test datasets, with 623,054, 77,866 and 78,592 samples respectively. Only the training and validation sets are used during training, fine-tuning parameters and model selection. This dataset includes a diversity of waveform characteristics. It also includes the different types of seismometers in Northern California Seismic Network. The proportion of each type in the dataset is shown in Figure 2. We select both high and low signal-to-noise ratio (SNR) recordings (Figure 3). The SNR is calculated by the ratio of standard deviations of the five seconds following and the five seconds preceding the P arrival. The complexity of this dataset makes it challenging for automatic phase picking, but it provides a more comprehensive performance evaluation.

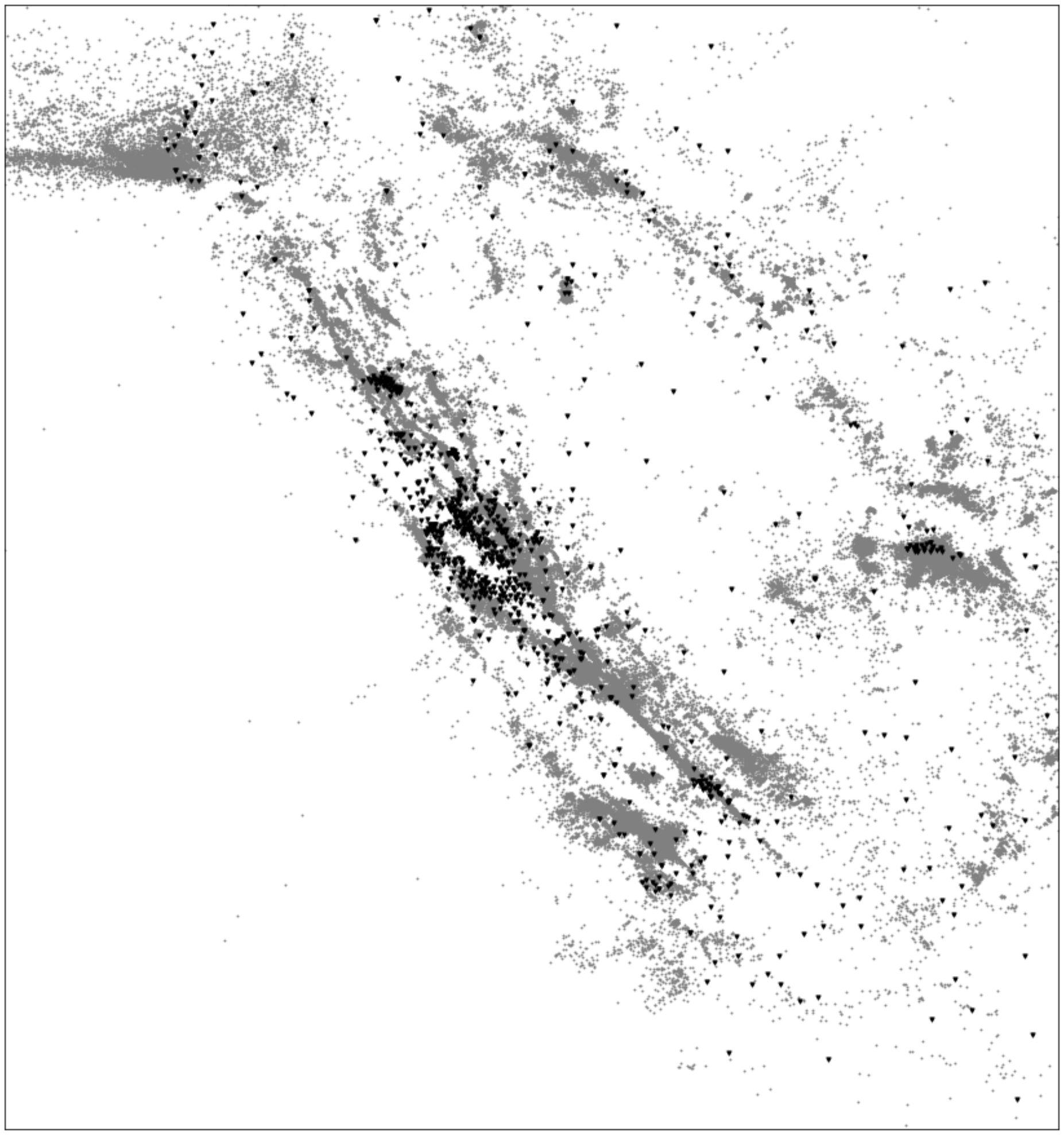

**Figure 1.** The locations of 234,117 earthquakes (grey points) and 889 seismic stations (black triangles) in the Northern California Earthquake Catalog.

We apply minimal data preprocessing to the training data. We normalize each component waveform by removing its mean and dividing it by the standard deviation (Figure 4(a-c)). All data are sampled to 100 Hz, which is the most common sampling rate in the dataset. The P/S arrival times are converted to P/S probability distributions (Figure 4(d)). That is, the arrival time data are represented probabilistically using a Gaussian distribution with zero mean and a standard deviation of 0.1 s. The arrival-times in the training dataset contain errors and biases. Representing them probabilistically allows the algorithm to reduce the influence of this uncertainty. It also helps accelerate convergence because it increases the amount of information on P ans S picks relative to noise, in much the same way that the extra information in cross correlation of waveforms improves arrival time measurements for similar earthquakes.

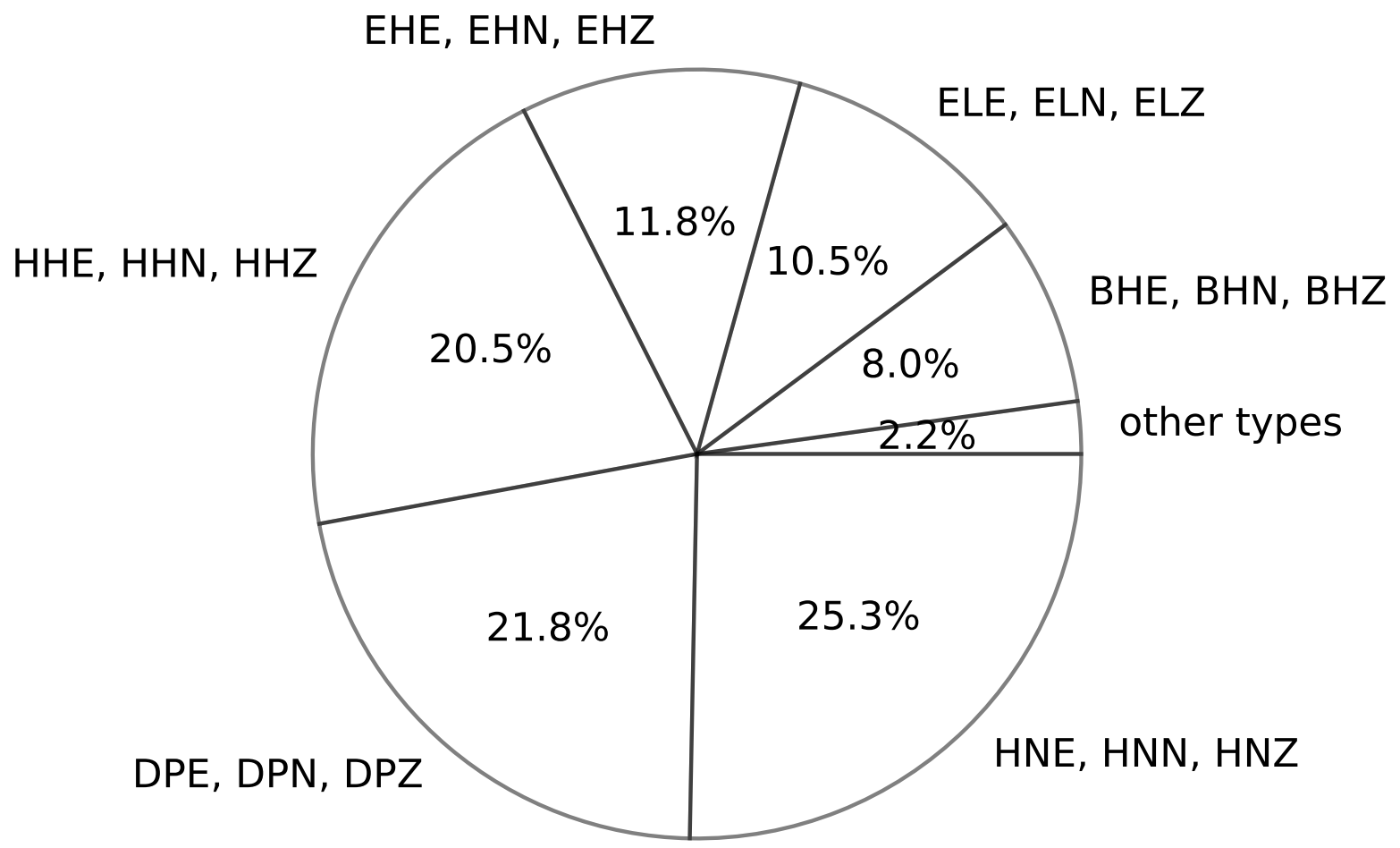

**Figure 2.** The proportion of different instrument type in the dataset. The first letter is the band code: H: high broad band, D: very very short period, E: short period. The second letter is the instrument code: N: accelerometer, P: very short period seismometer, H: high gain seismometer, L: low gain seismometer. The third letter is the orientation code: E: east-west direction, N: north-south direction, Z: vertical direction.

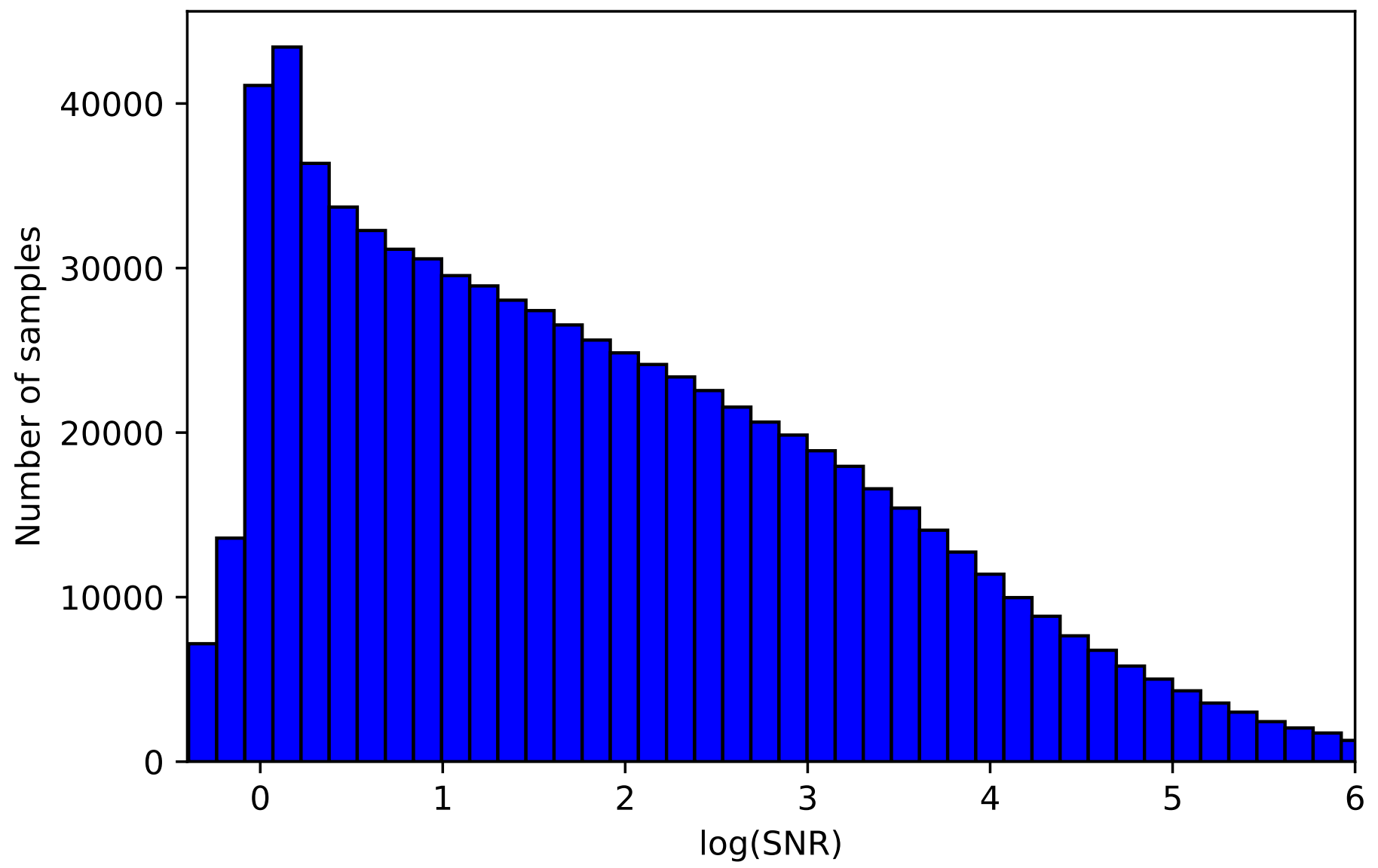

**Figure 3.** Signal-to-noise ratio (SNR) distribution. The SNR is calculated by the ratio of standard deviations of two five-seconds windows following and preceding the P arrival. Note that we purposely include large amounts of low SNR data to improve arrival time measurement for small events.

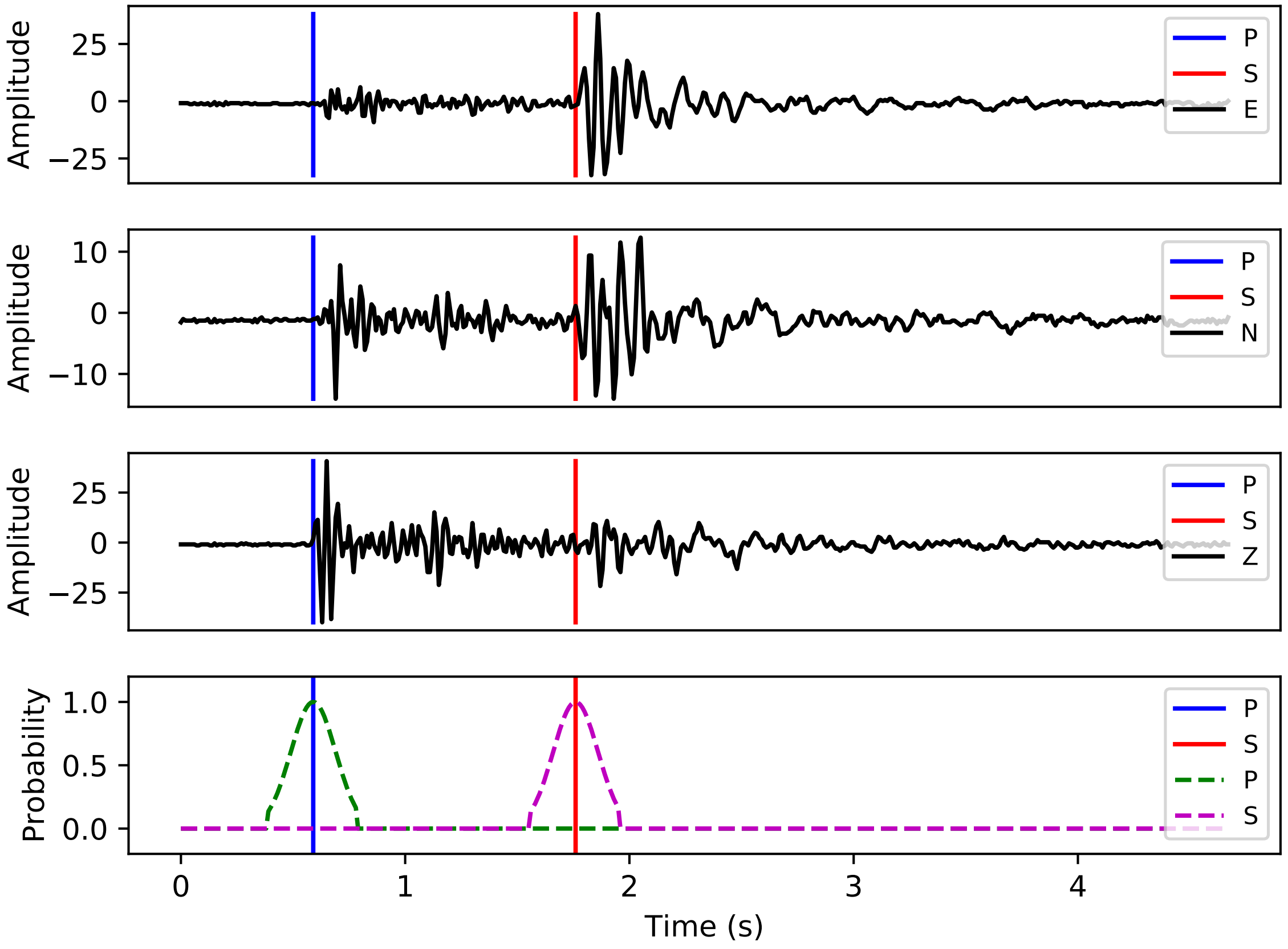

**Figure 4.** A sample from the dataset. (a) - (c) Seismograms of the "ENZ" (East, North, Vertical) components. The blue and red vertical lines are the manually picked P and S arrival times. (d) The converted probability distribution for P and S pickers. The shape is a truncated Gaussian distribution with mean ($\mu = 0s$) and standard deviation ($\sigma = 0.1s$).

## 3 METHOD

The architecture of PhaseNet (Figure 5) is modified from U-net (Ronneberger et al. 2015) to deal with 1-D time series data. U-net is a deep neural network approach used in biomedical image processing, that seeks to localize properties in an image. The mapping to our problem is to localize the properties of our time series into three classes: P picks, S picks, and noise. The inputs are three-component seismograms of known earthquakes. The outputs are probability distributions of P wave, S wave, and noise. The softmax normalized exponential function is used to set probabilities in the last layer:

$$q_i(x) = \frac{e^{z_i(x)}}{\sum_{k=1}^{3} e^{z_k(x)}}$$

where $i = 1, 2, 3$ represents noise, P and S categories. $z(x)$ are the unscaled values of the last layer. The loss function is defined using cross entropy between the true probability distribution $(p(x))$ and predicted distribution $(q(x))$:

$$H(p, q) = -\sum_{i=1}^{3} \sum_{x} p_i(x) \log q_i(x),$$

which measures the divergence between the two probability distributions.

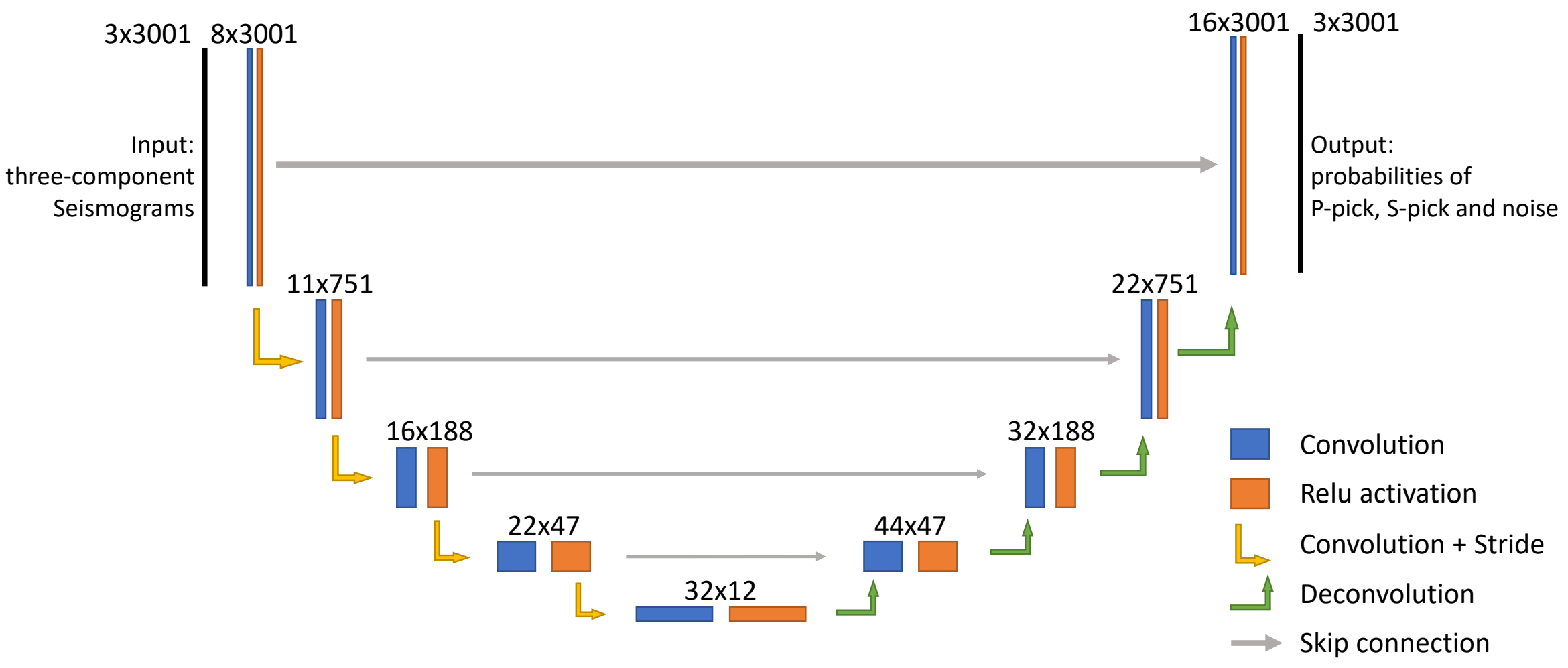

**Figure 5.** The network architecture

The input seismic data go through four down-sampling stages and four up-sampling stages. Inside each stage, we apply convolution and rectified linear unit (ReLU) activation. The down-sampling process is designed to extract and shrink the useful information from raw seismic data to a few neurons, so each neuron in the last layer makes up a broadly receptive window. The up-sampling process expands and converts this information into probability distributions of P wave, S wave and noise for each time point. A skip connection at each depth directly concatenates the left output to the right layer without going through the deeper layer. This should help improve convergence during training (Ronneberger et al. 2015). The convolution size is set to 7 points and the stride step for down-sampling is set to 4 points. The size of each layer is shown in Figure 5. The P and S first arrival times are extracted from the peaks of output probability distributions.

| Evaluation Indicator | Phase | PhaseNet | AR picker |
|---|---|---|---|
| Precision | P | 0.939 | 0.558 |
| | S | 0.853 | 0.195 |
| Recall | P | 0.857 | 0.558 |
| | S | 0.755 | 0.144 |
| F1 score | P | 0.896 | 0.558 |
| | S | 0.801 | 0.165 |
| $\mu(\Delta t)(ms)$ | P | 2.068 | 11.647 |
| | S | 3.311 | 27.496 |
| $\sigma(\Delta t)(ms)$ | P | 51.530 | 83.991 |
| | S | 82.858 | 181.027 |

**Table 1.** Evaluation metrics on the test dataset. Pickers with residuals ($\Delta t < 0.1s$) are counted as correct. The mean ($\mu(\Delta t)$) and standard deviation ($\sigma(\Delta t)$) are calculated on residuals ($\Delta t < 0.5s$) whose distributions are shown in Figure 6

## 4 EXPERIMENTS

We have chosen the evaluation metrics: precision, recall, F1 score, mean ($\mu$) and standard deviation ($\sigma$) of time residuals ($\Delta t$) between our picks and ground truth to test the performance of PhaseNet (Powers 2011). Precision, recall and F1 are defined as:

$$precision: P = \frac{T_p}{T_p + F_p}$$

$$recall: R = \frac{T_p}{T_p + F_n}$$

$$F1 = 2\frac{P \times R}{P + R}$$

where $T_p$ is the number of true positives, $F_p$ is the number of false positives, and $F_n$ is the number of false negatives. Peak probabilities above 0.5 are counted as positive picks. Arrival-time residuals that are less than 0.1s ($\Delta t < 0.1s$) are counted as true positives. Picks with

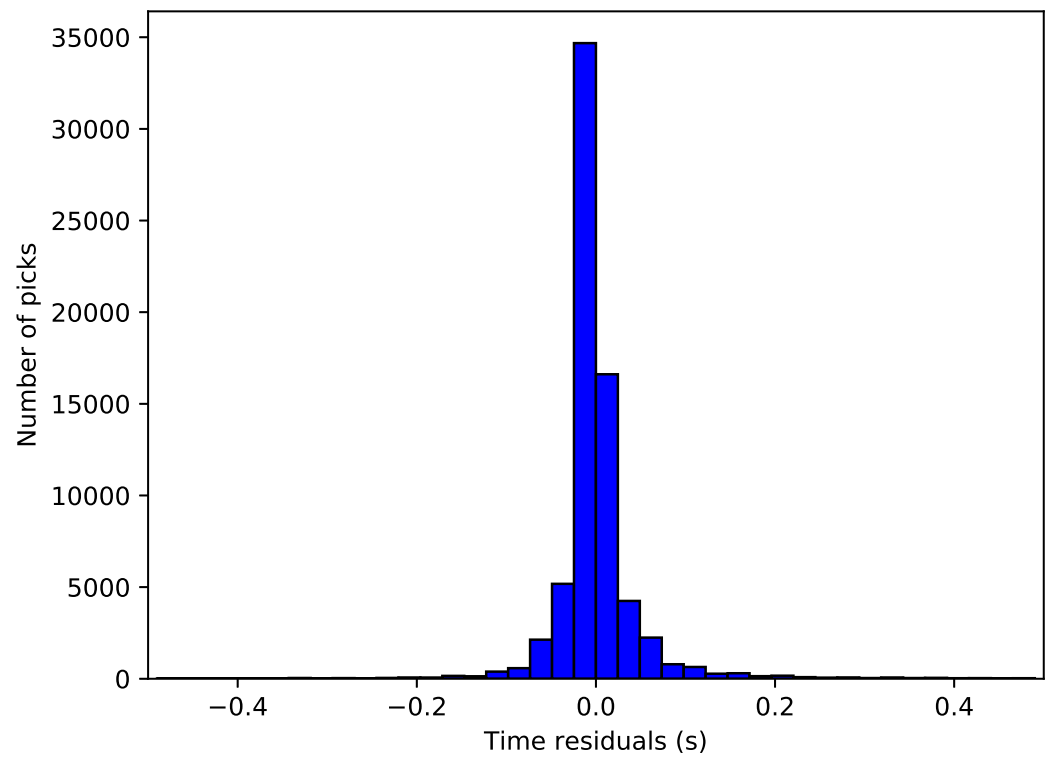

(a) P picks of PhaseNet

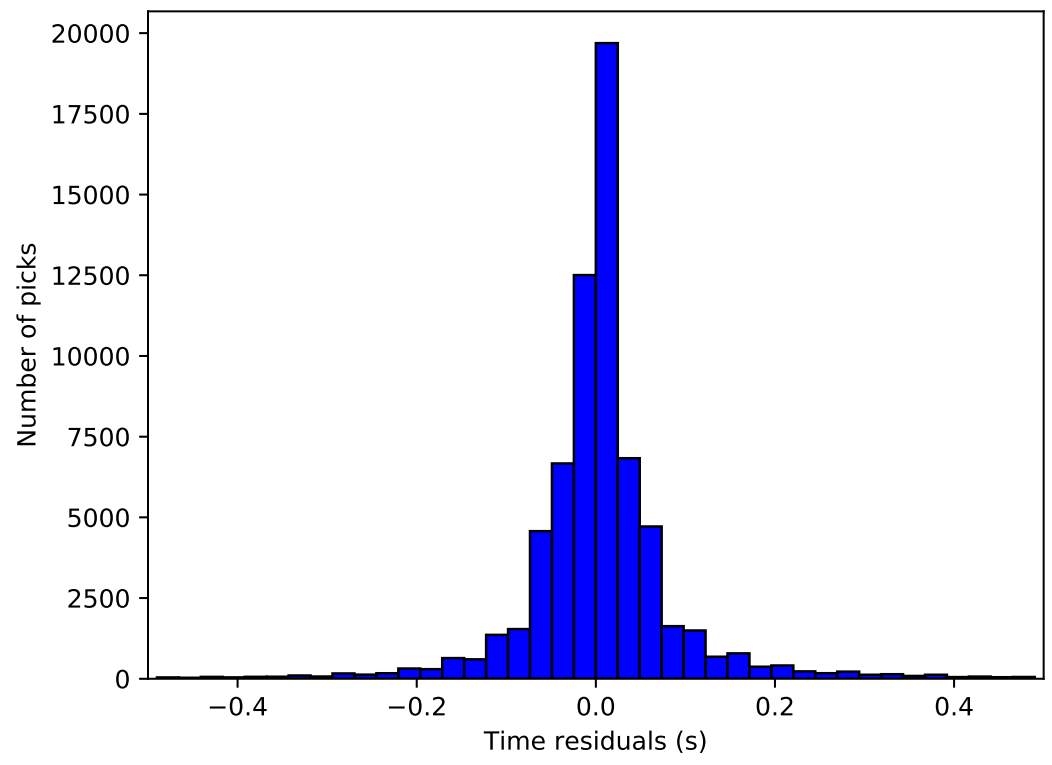

(b) S picks of PhaseNet

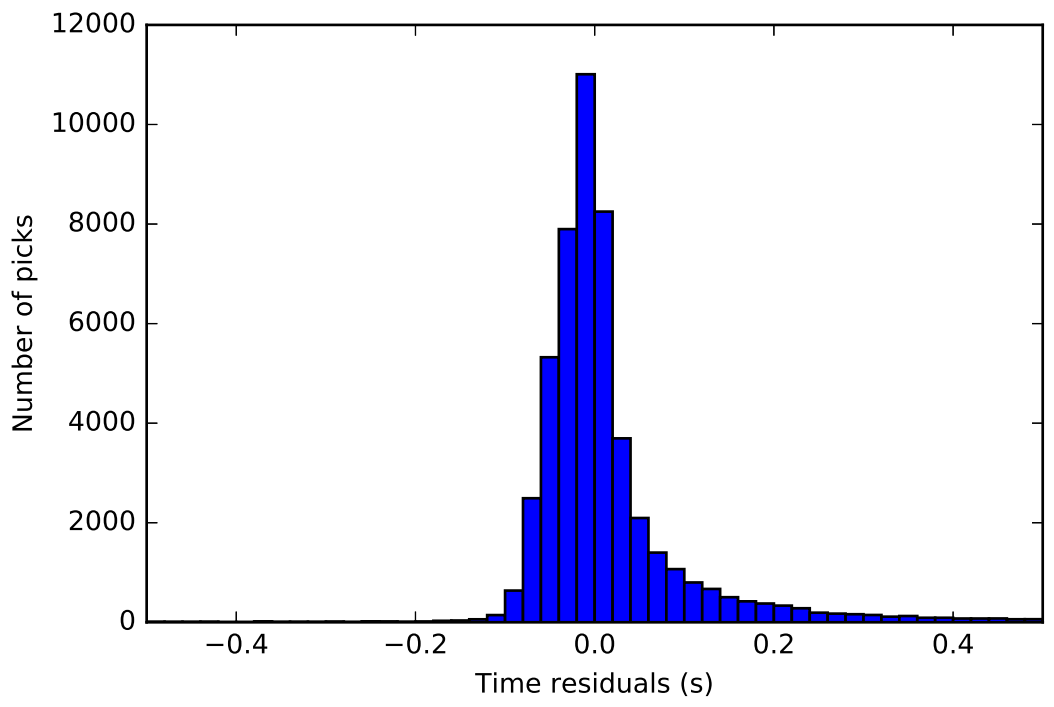

(c) P picks of AR picker

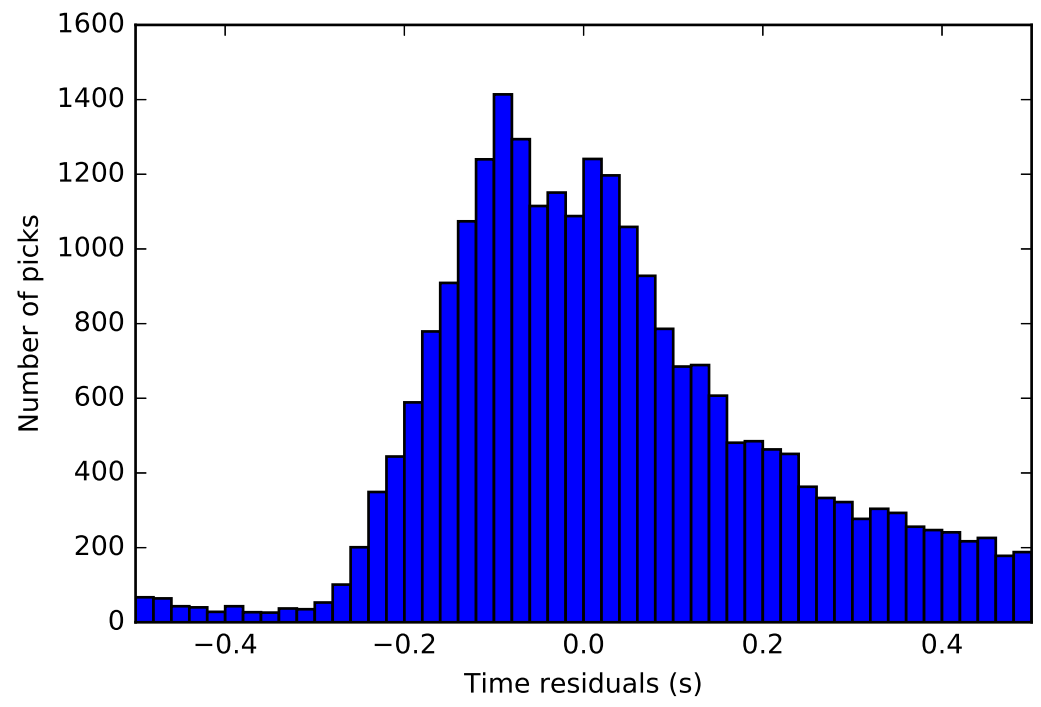

(d) S picks of AR picker

**Figure 6.** The distribution of residuals ($\Delta t$) of PhaseNet (upper panels) and AR picker (lower panels) on the test dataset

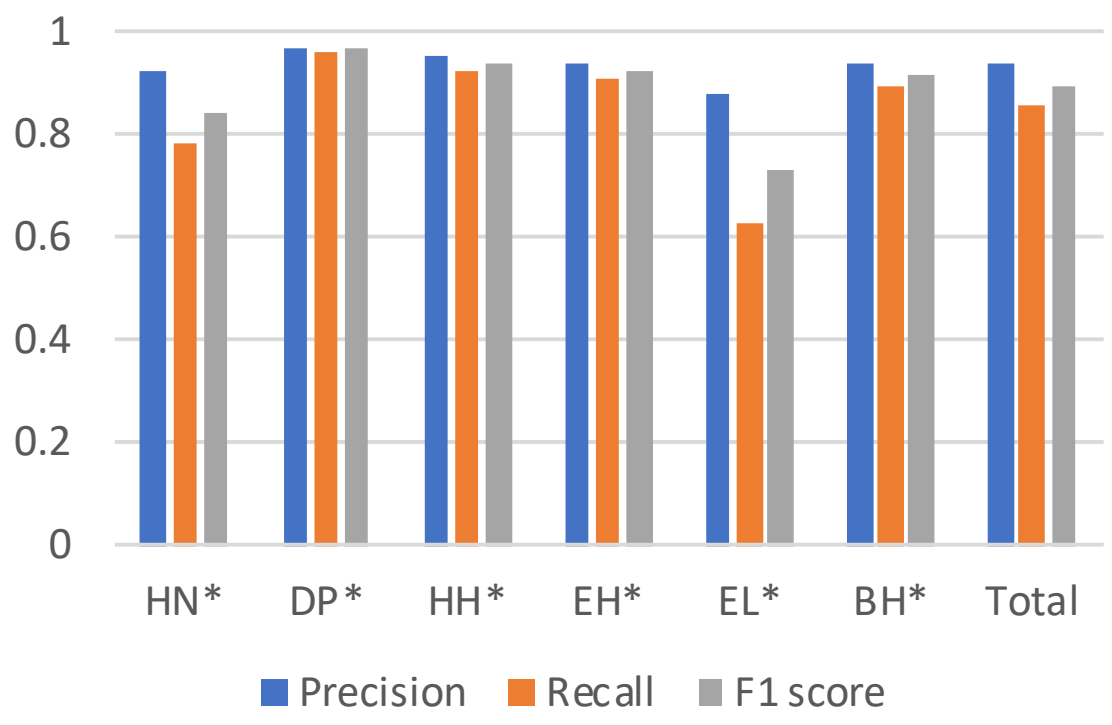

(a) P phase

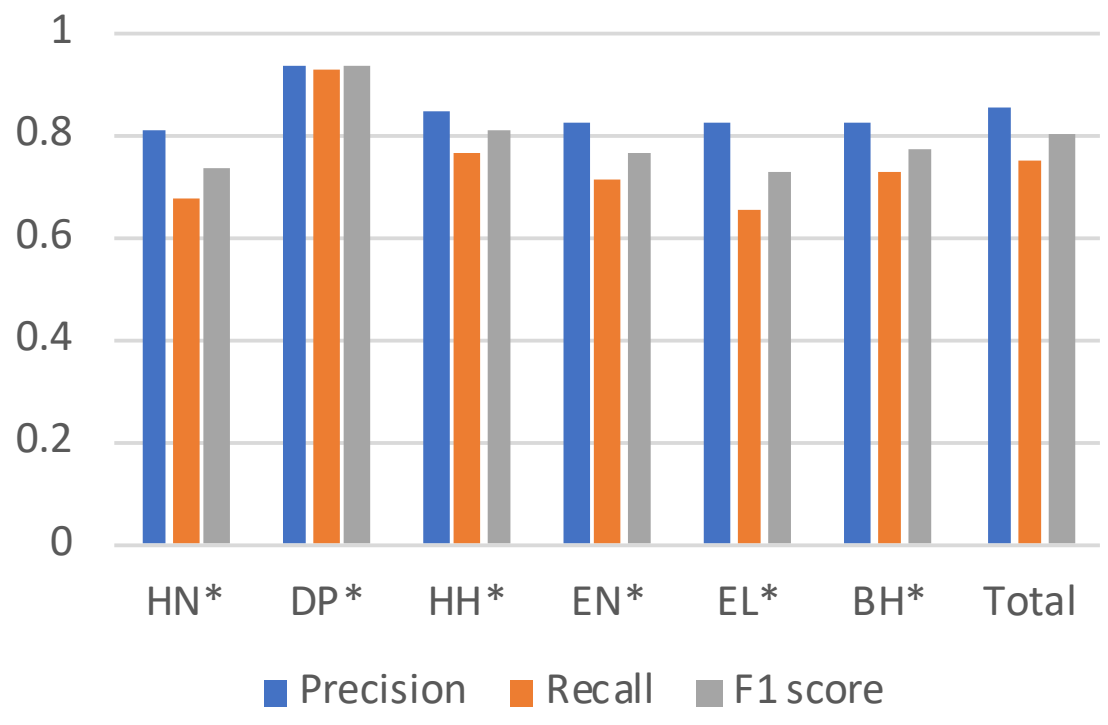

(b) S phase

**Figure 7.** Performances on different instrument types. (a) P picks. (b) S picks. The meaning of x-axis labels are the same as Figure 2. The "total" dataset is the same test dataset used in Table 1.

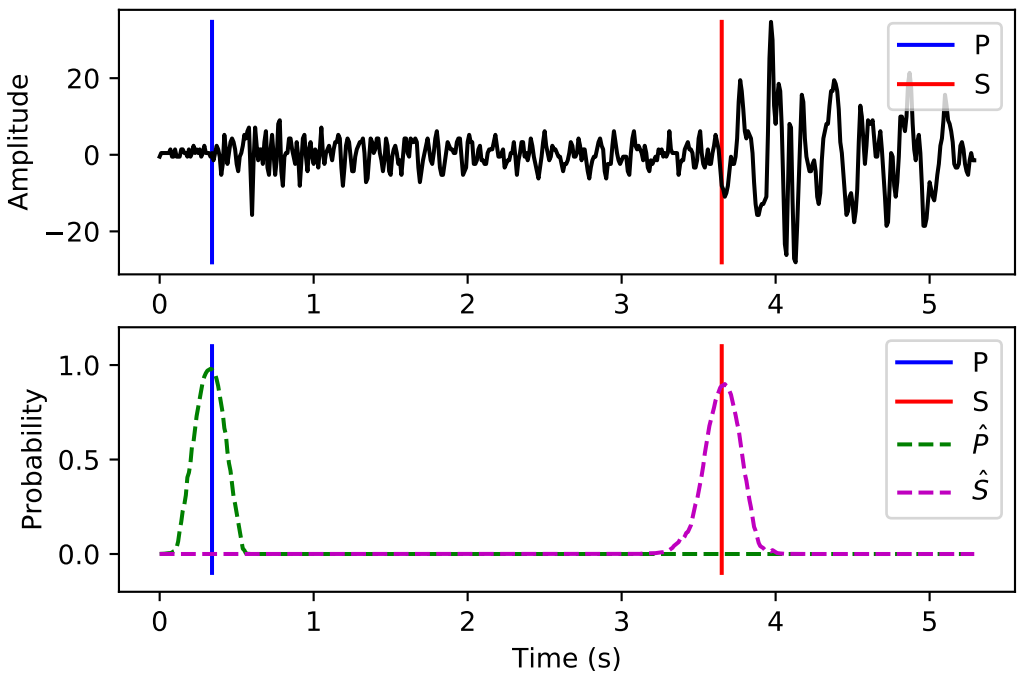

(a)

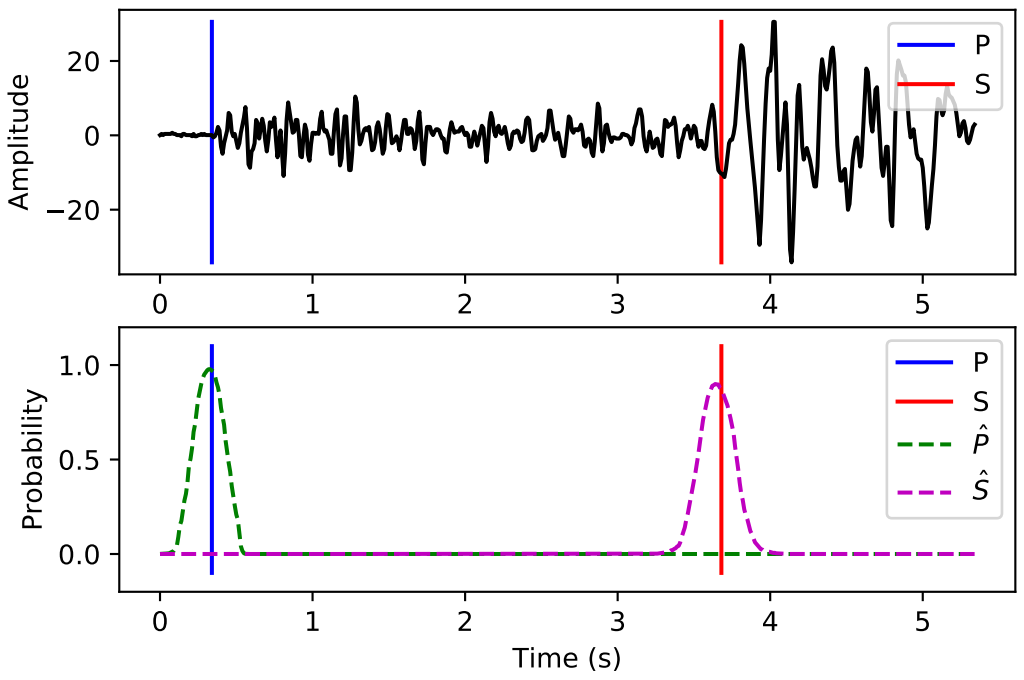

(b)

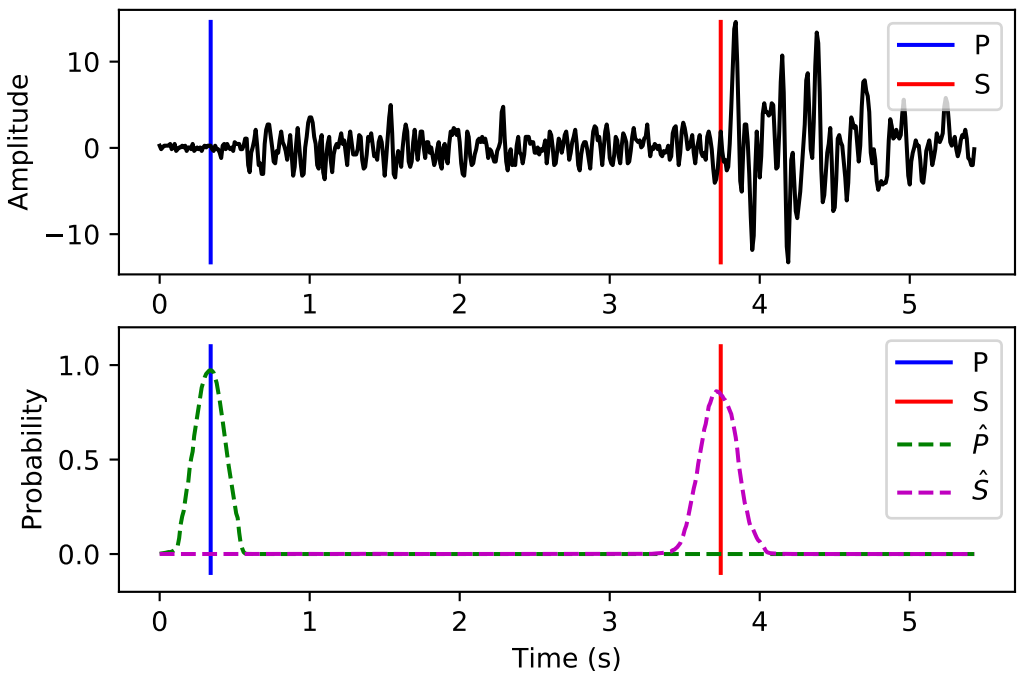

(c)

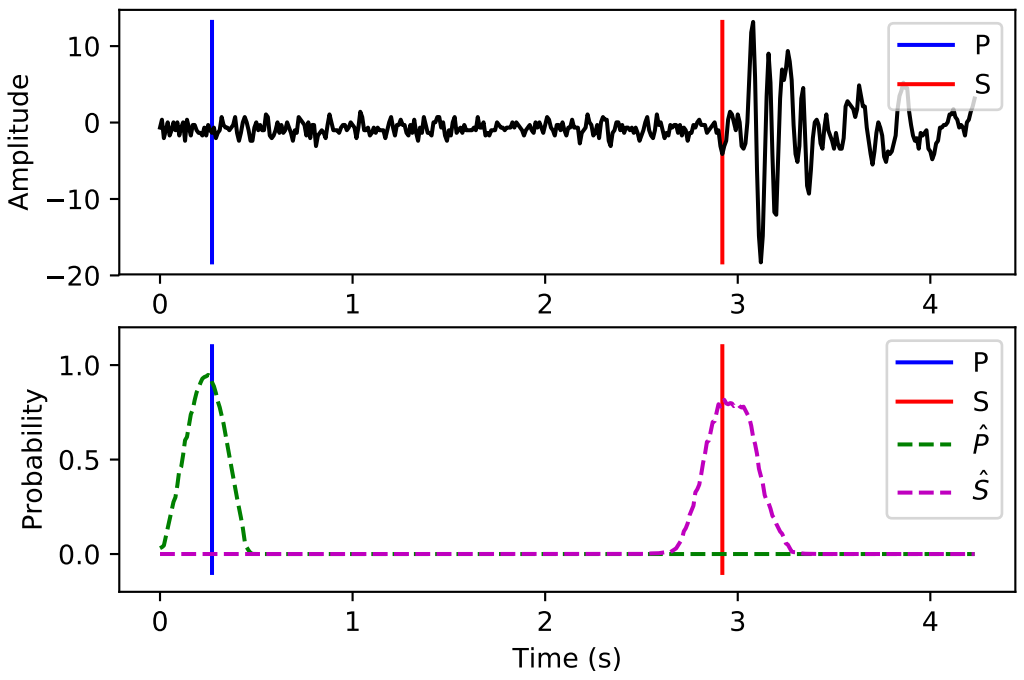

(d)

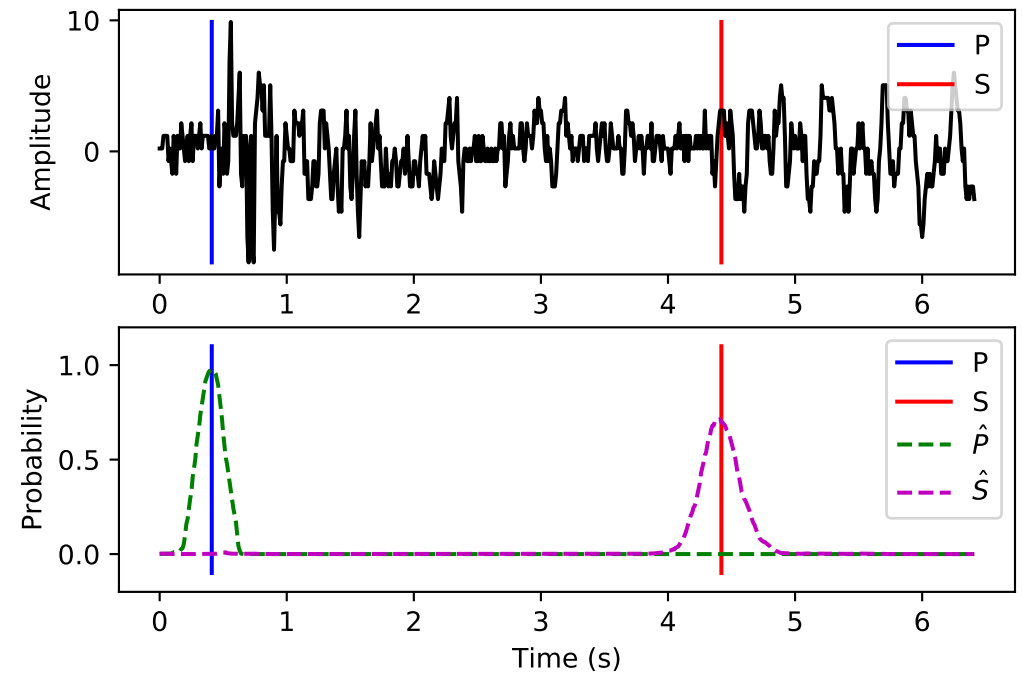

(e)

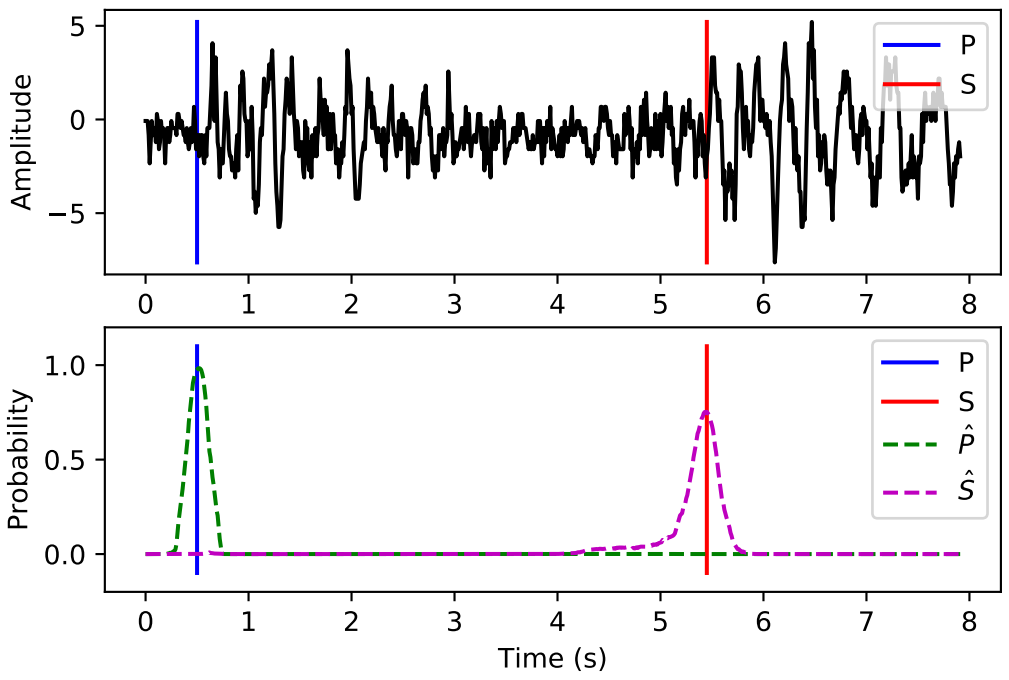

(f)

**Figure 8.** Examples of good pickers ($\Delta t < 0.1s$) in the test dataset. The upper parts of (a) - (f) sub-figures are the vertical components of seismograms. The lower parts are the predicted probability distributions of P wave ($\hat{P}$) and S wave ($\hat{S}$). The blue and red vertical lines are the P and S arrival times picked by analysts. While all three components are used in PhaseNet, in this and subsequent figures, only the vertical component is shown.

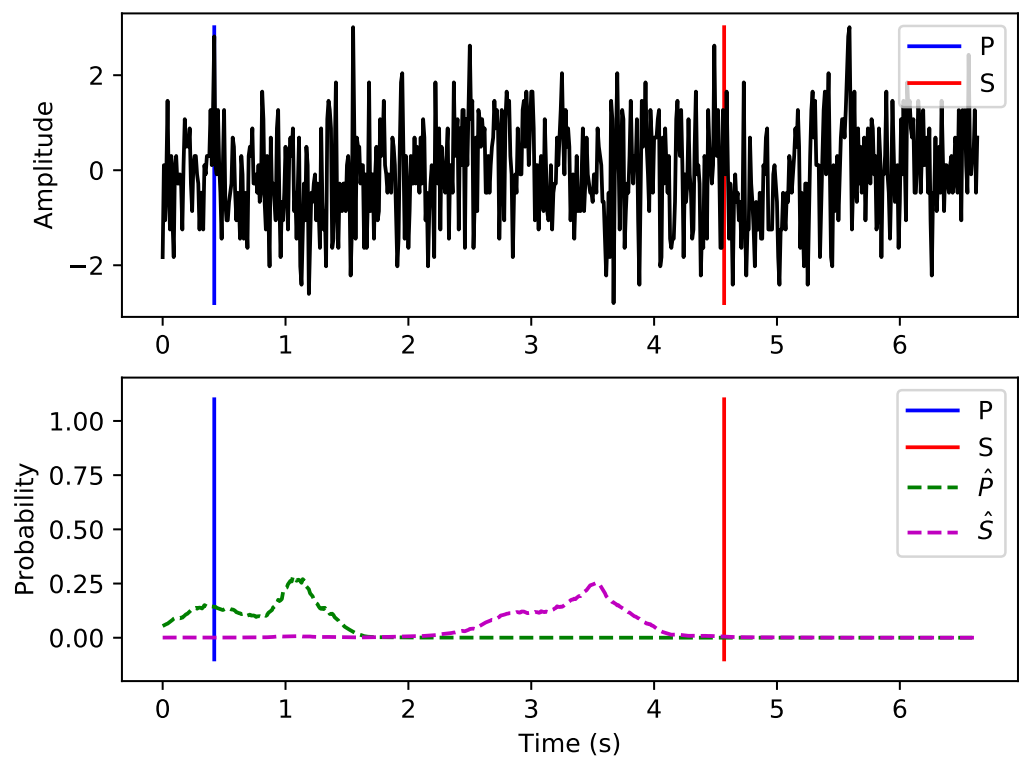

(a)

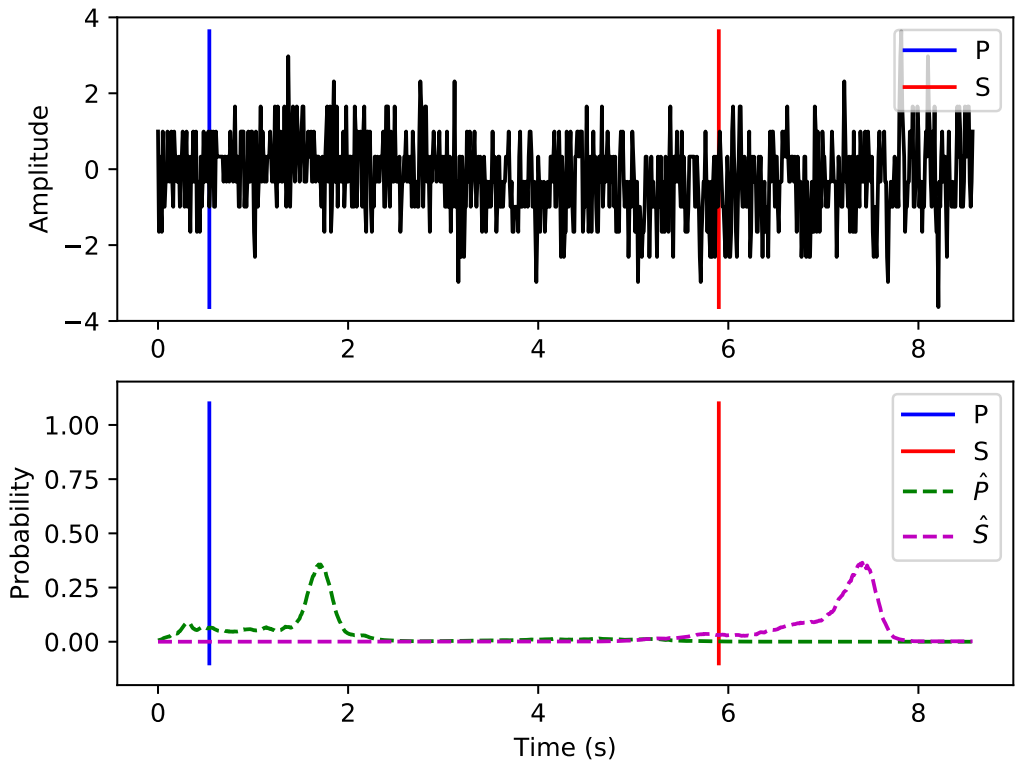

(b)

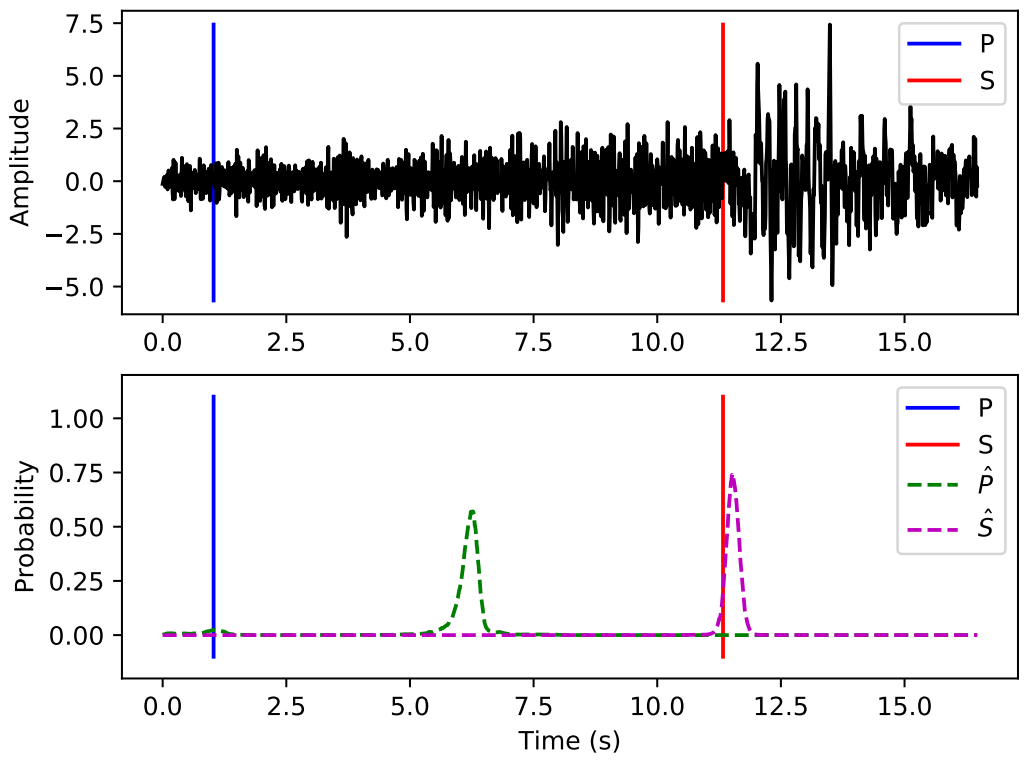

(c)

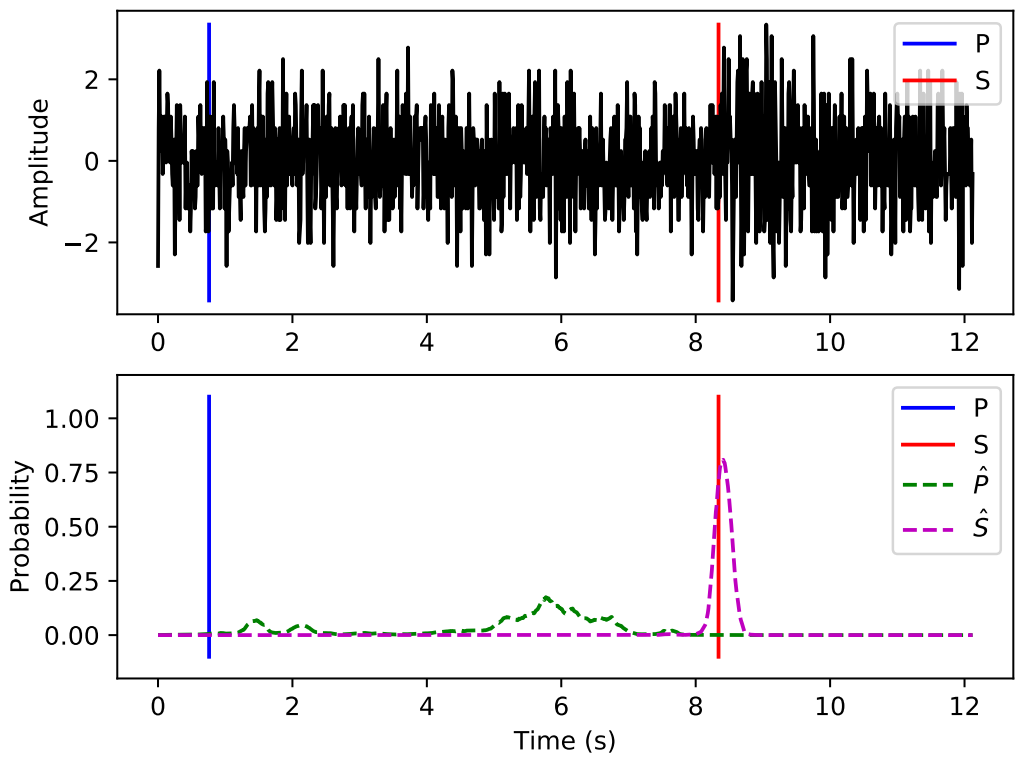

(d)

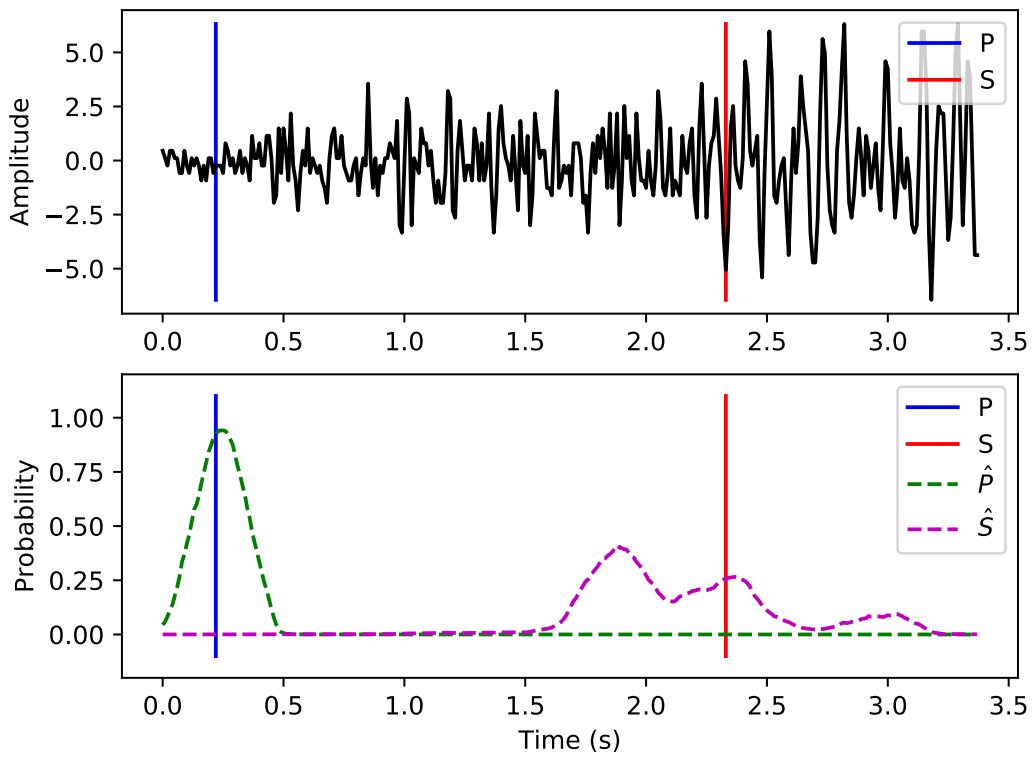

(e)

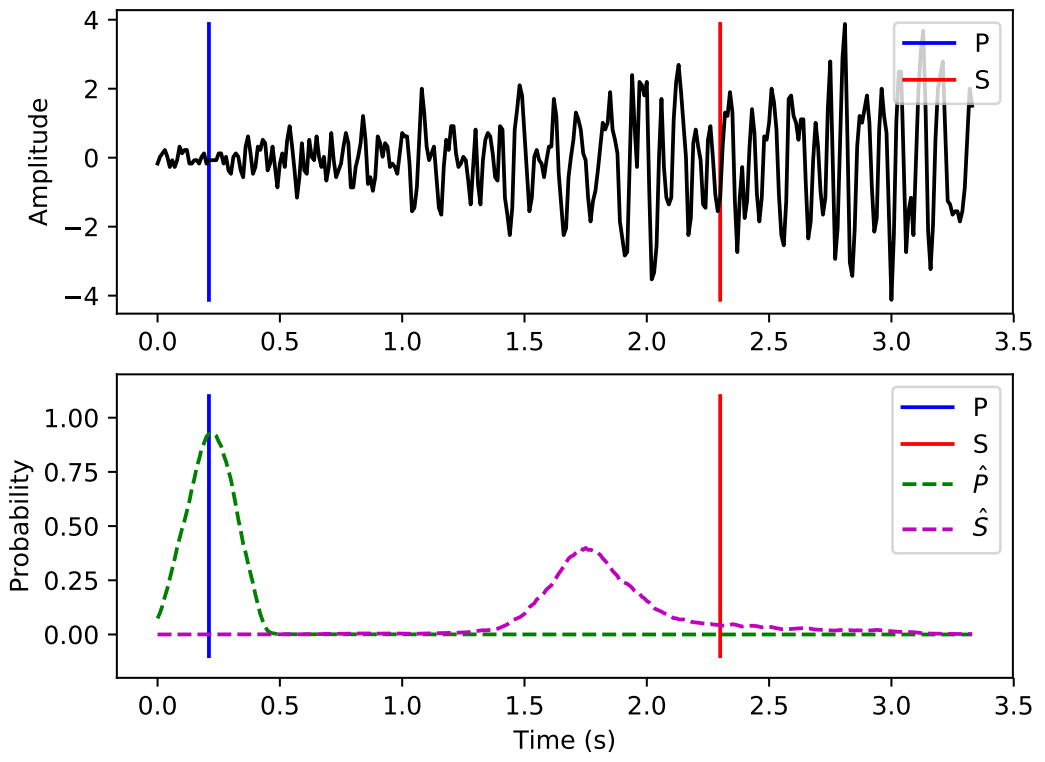

(f)

**Figure 9.** Examples of bad picks in the test dataset. (a, b) are examples of no P or S picks predicted. (c, d) are examples of bad P picks. (e, f) are examples of bad S picks.

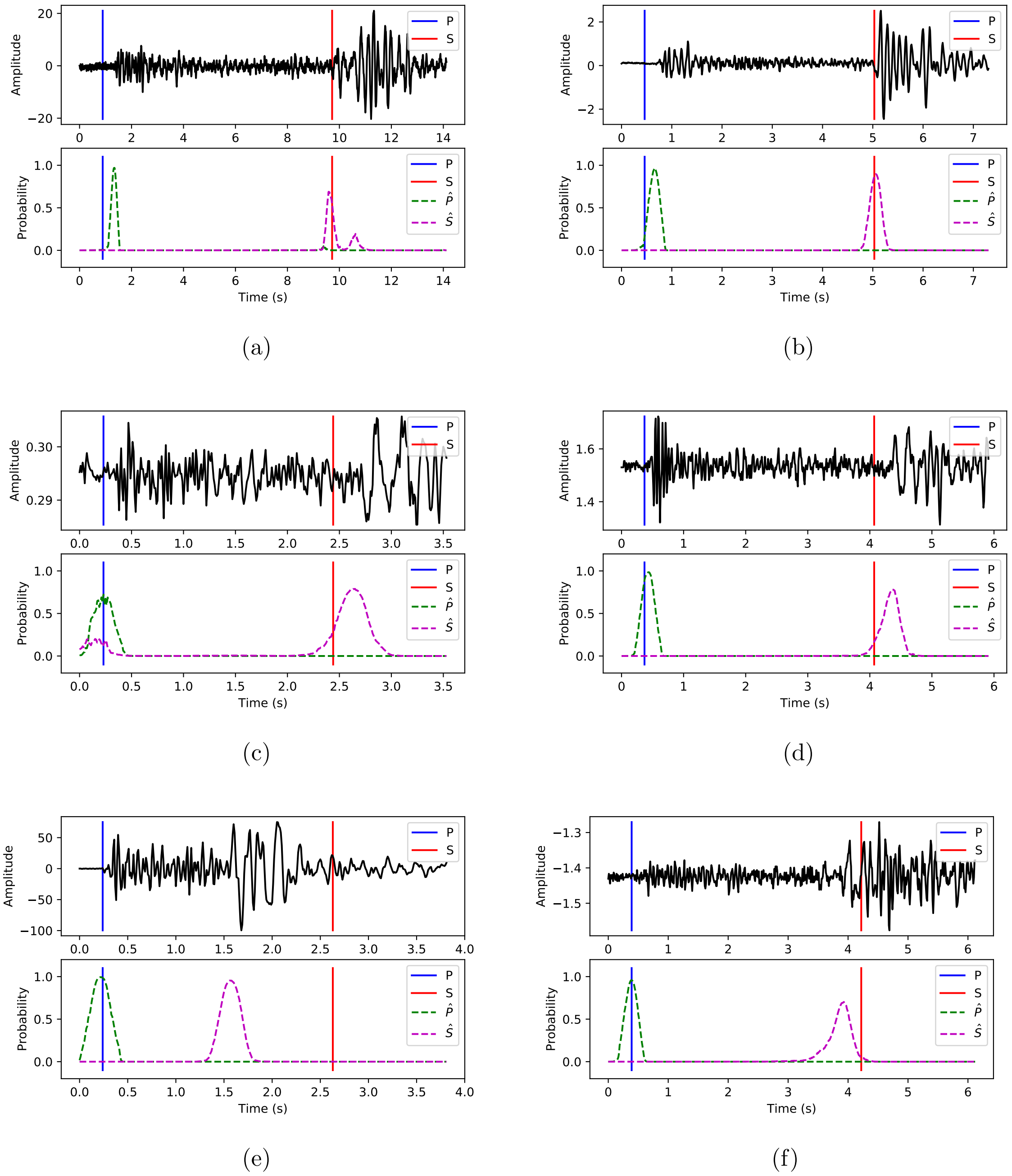

**Figure 10.** Examples where manual picks may be not accurate in the test dataset. (a, b) are ambiguous P picks. (c) - (f) are ambiguous S picks.

larger residuals are counted as false positives. We compare our results with those obtained by the open-source "AR picker" (Akazawa 2004) implemented in Obspy (Beyreuther et al. 2010). The results of both PhaseNet and AR picker are shown in Table 1. For our data set, our method achieved significant improvements, particularly for the S waves. Because S waves emerge from the scattered waves of the P coda, picking S arrivals is more challenging for automatic methods.

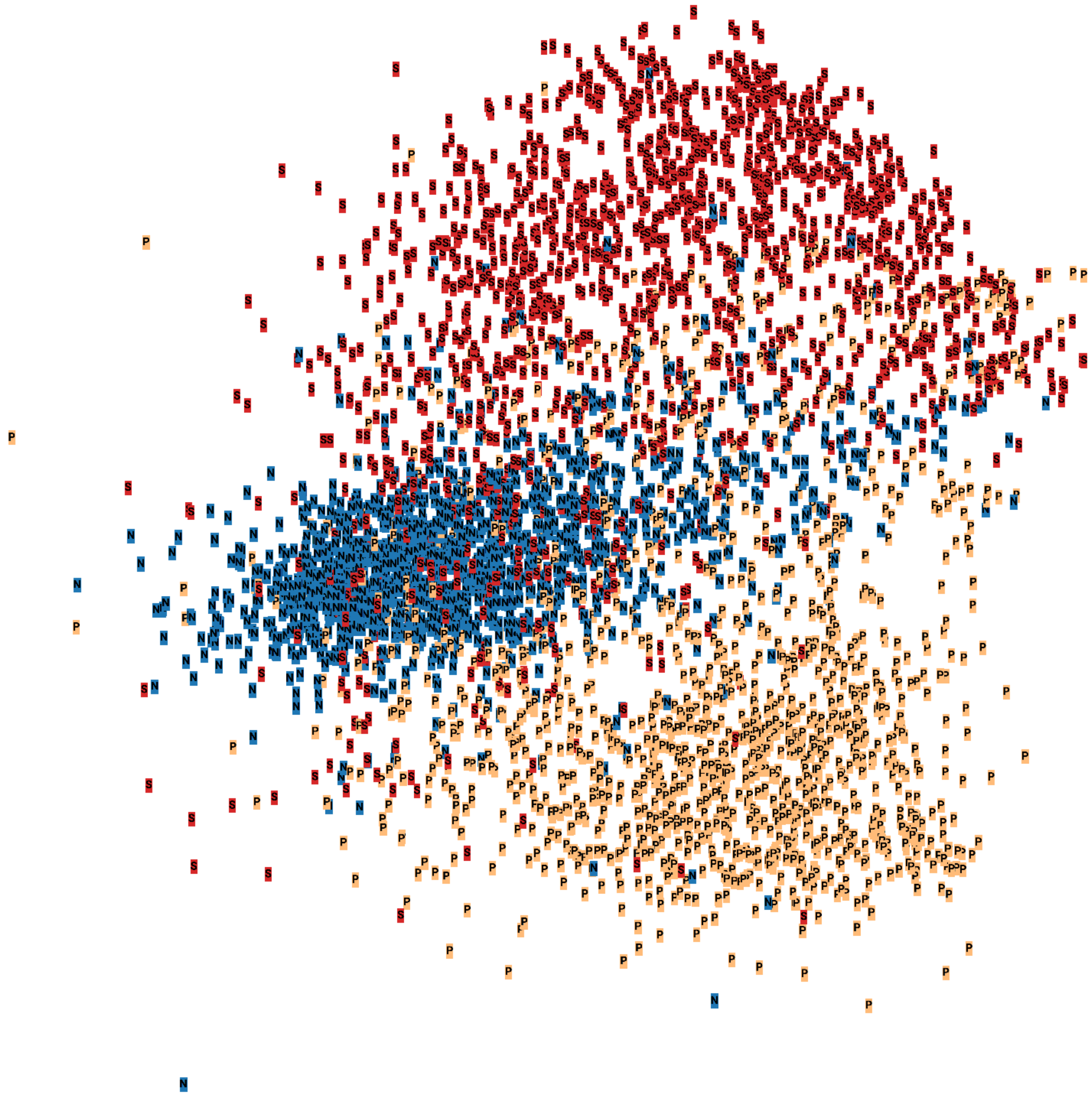

**Figure 11.** PCA visualization of weights in the deepest layer. The red, yellow and blue dots represent input data with P picks, S picks or only noise.

Figure 6 shows the distribution of time residuals between the automated and human-labeled P and S picks. The residual distributions of the P wave picks are much narrower than for the S wave picks, which is consistent with the fact that P wave arrivals are expected to be clearer and hence easier to pick. The residual distributions of both P and S wave arrivals for PhaseNet are distinctly narrower and do not have obvious biases compared with the results from the AR picker.

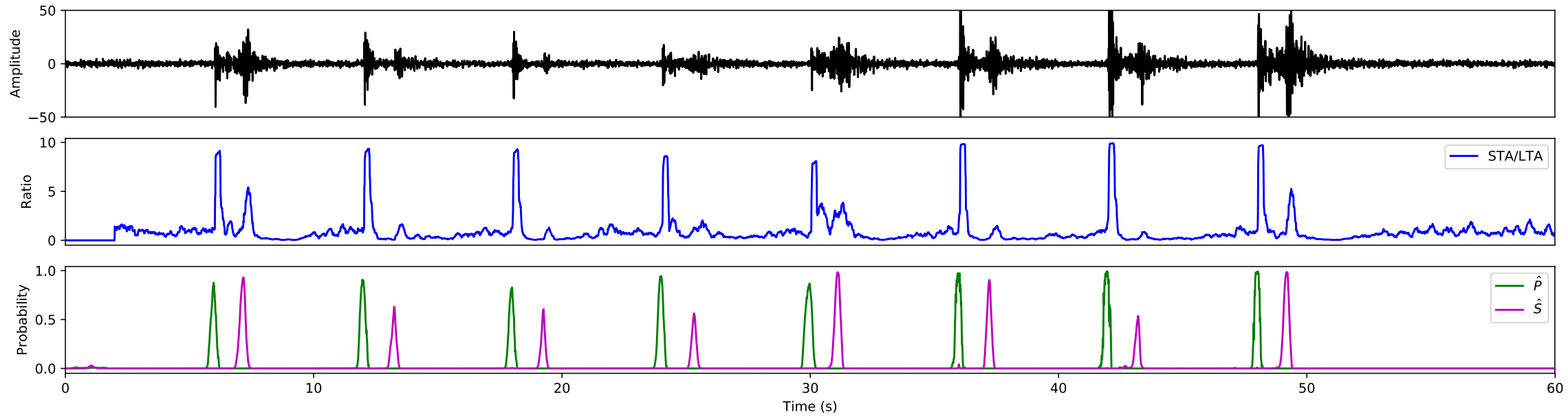

**Figure 12.** Synthetic continuous seismic waveforms. (a) waveform of vertical component. (b) output of basic STA/LTA in Obspy. (c) output of PhaseNet. The continuous data is created by stacking waveforms of eight events. The first-arrival-time interval between adjacent events is six seconds. The STA/LTA method runs on vertical component. The PhaseNet runs on three components.

Figure 7 shows performances of PhaseNet on different instrument types. The same model, which is trained on all instrument types, is used for testing here; however the test set is divided based on each instrument type. Without changing any parameters or thresholds, the performance of PhaseNet is robust on different instruments. Despite the waveform differences between short period and broad band, high gain and low gain, accelerometer and seismometer, PhaseNet learns the common features needed to detect P and S phases and picks the correct arrival times.

It is instructive to look at a handful of representative results. Figure 8 shows good examples from the test dataset. The peaks of the predicted distributions accurately align with the true P and S picks. Figures 8(c, d) and Figures 8(e, f) show more ambiguous cases with no clear abrupt changes around the P or S picks. PhaseNet can still predict the correct arrival times with high confidence. Figure 9 shows some apparently failed cases. The P and S first arrivals are harder to distinguish and the waveforms are more noisy and complex than those in Figure 8. Figure 10 shows some interesting cases where the P or S arrival times picked by analysts may be incorrect. The predictions of the neural networks appear more reasonable and consistent. Because there are subjective factors in seismic-phase picking, analysts may use different criteria to pick arrivals. Picks by the same analysts may also differ at different times.

To analyze the representations that PhaseNet has learned, we train another model without the skip connection and apply PCA (Principal Component Analysis) analysis to the neural weights of the deepest layer (Figure 5). The neural network condenses the knowledge from high dimensional raw waveforms into a few parameters in the deepest layer, which means that these low dimensional neural weights should contain the information needed to determine P vs. S arrivals. We feed in seismic data with P picks, S picks or only noise, and record the corresponding vectors in the deepest layer. The PCA visualization (Figure 11) shows that these condensed vectors group to different regions for P, S, and noise. This demonstrates that the neural network has learned to extract the characteristic features of P waves, S waves, and noise from the raw data and capture them in the condensed neural weights in the deepest layer.

PhaseNet predicts the probability distributions of P and S picks for every data point in the time series, so it may be applied to continuous data for earthquake detection. We have created continuous seismic data by stacking waveforms of eight different events (Figure 12). These events are shifted to make the arrival-time interval between adjacent events equal to six seconds. We have applied both basic STA/LTA in Obspy and our PhaseNet method on this sequence. The lengths of the short and long window of STA/LTA method are chosen as 0.2s and 2s respectively. The output sequences in Figure 12 show that PhaseNet produces similar spikes as STA/LTA methods, which are commonly used for earthquake detection; however, PhaseNet can also differentiate between P and S arrivals. This information may also used to reduce false detections, because events with both P and S picks are more likely to be a true earthquakes compared with the undifferentiated spikes reported by STA/LTA.

## 5 DISCUSSION

We have shown that PhaseNet can detect and pick P and S arrivals effectively within known earthquake waveforms. The F1 score provides a balanced assessment of algorithm performance in both precision and recall. PhaseNet achieves an F1 score of 0.896 for P arrivals and 0.801 for S arrivals, which is substantially better than the AR picker (0.558 for P arrivals and 0.165 for S arrivals). We have chosen a strict threshold for true positive ($\Delta t < 0.1s$) during evaluation. If we were to relax this standard, the F1 score would be even higher. Our method differs from that proposed by Ross & Ben-Zion (2014), because PhaseNet does not explicitly use polarization analysis to separate P from S waves. PhaseNet automatically learns features, which might implicitly include polarization, to distinguish P from S waves. We find that the improvement S wave picks is more significant than the improvement to P phase picks, which suggests that the features learned from data are more effective than manually defined features.

The STA/LTA method is based on detecting a sudden change in waveform amplitude. But the S phase is always contaminated by P coda, which degrades the ability of the STA/LTA ratio to make an accurate S pick. PhaseNet has an advantage here in that it can learn features other than amplitude both to detect S waves and to differentiate between P and S waves. Figure 13 shows examples of PhaseNet applied to clipped waveforms. Although the amplitude is strongly clipped, PhaseNet is still able to pick S arrivals successfully.

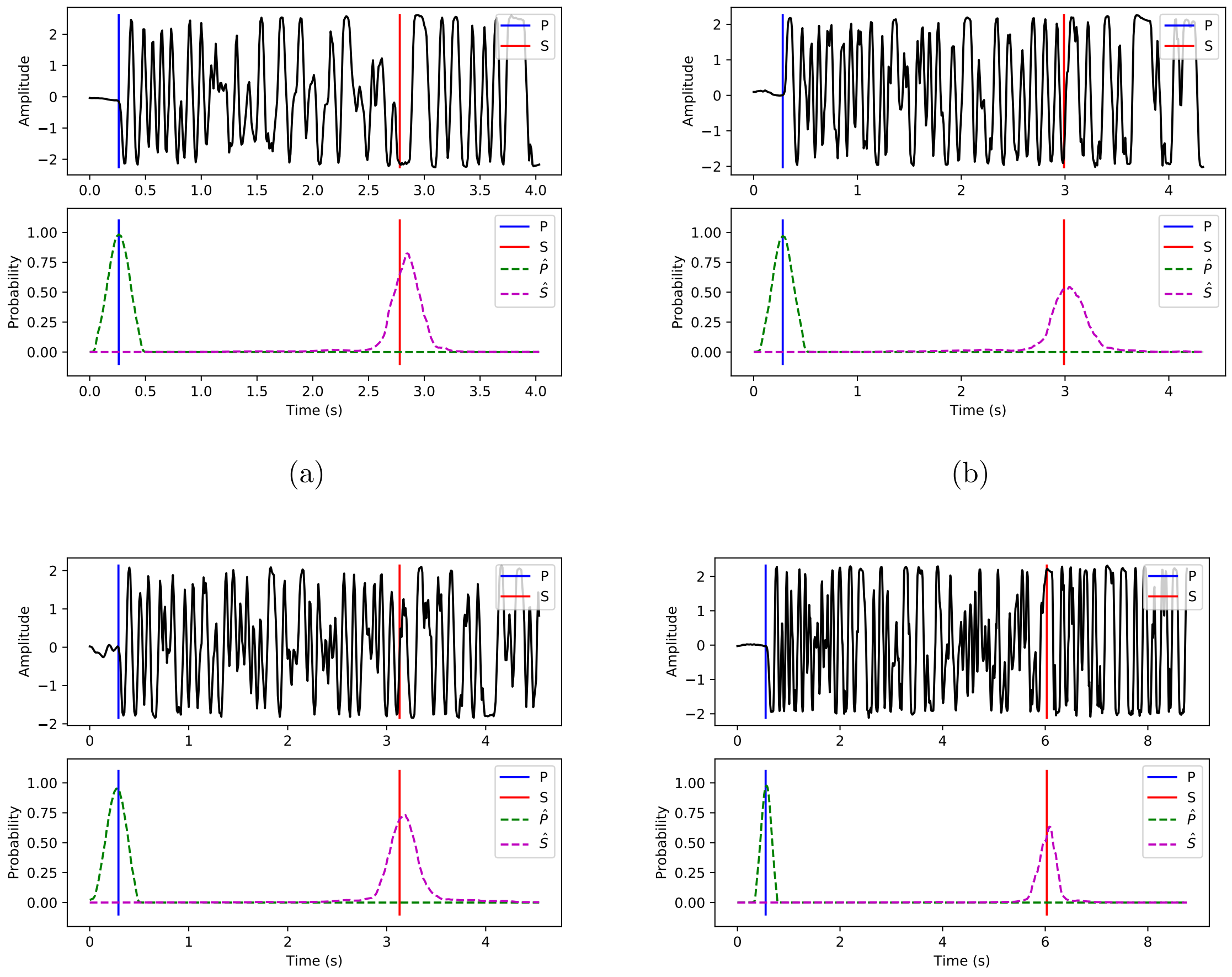

(a) (b)

(c) (d)

**Figure 13.** Examples of amplitude clipped waveforms.

We have not pre-processed the data with denoising techniques such as band-pass filtering. As a result, our dataset contains a number of low signal-to-noise ratio data. We apply the AR picker after pre-processing the data with a band-pass filter of 0.1Hz - 30Hz. Without filtering, its performance would be substantially degraded. PhaseNet does not require this pre-processing because it not only learns the characteristics of P and S waves, but it also learns what kind of data is noise. This means that it will still work reliably with noisy data, and to the extent that non-stationary noise is present in the training set, will be able to handle that too. Figure 14 shows several prediction results on low SNR data, for which it would be difficult for analysts to pick P and S arrivals. Despite these challenges, PhaseNet predicts accurate arrival-times at high probability. Figure 15 shows examples with strong low frequency background noise. PhaseNet can accurately pick both P and S phases without the need for filtering.

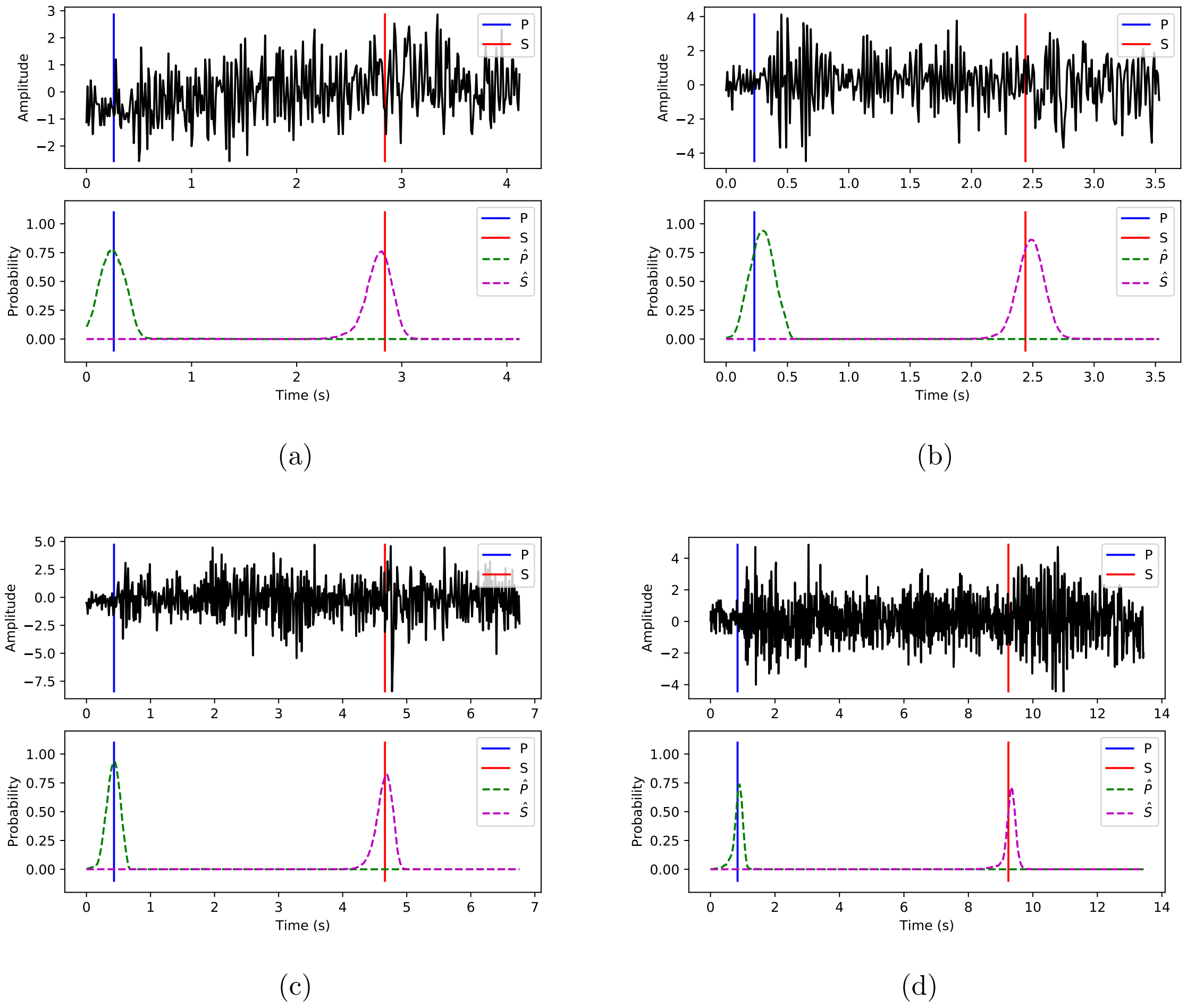

**Figure 14.** Examples of low SNR data

The STA/LTA method is sensitive to the threshold selected to determine P or S wave arrivals, and there is an inevitable trade-off between too high and too low a threshold. Moreover, it is prone to a delayed arrival-time if the threshold is set too high. Instead of an unbounded STA/LTA ratio, PhaseNet estimates a probability. We have set the threshold of probability to 0.5 for both P and S picks. Here too there is a trade-off, and tuning this threshold can further improve the performance, but the effect is not significant. Unlike STA/LTA, this threshold will not systematically bias arrival times, because this threshold is only used to decide if it is a pick. The accurate arrival time is measured from the peak of the probability distribution and does not depend strongly on this threshold.

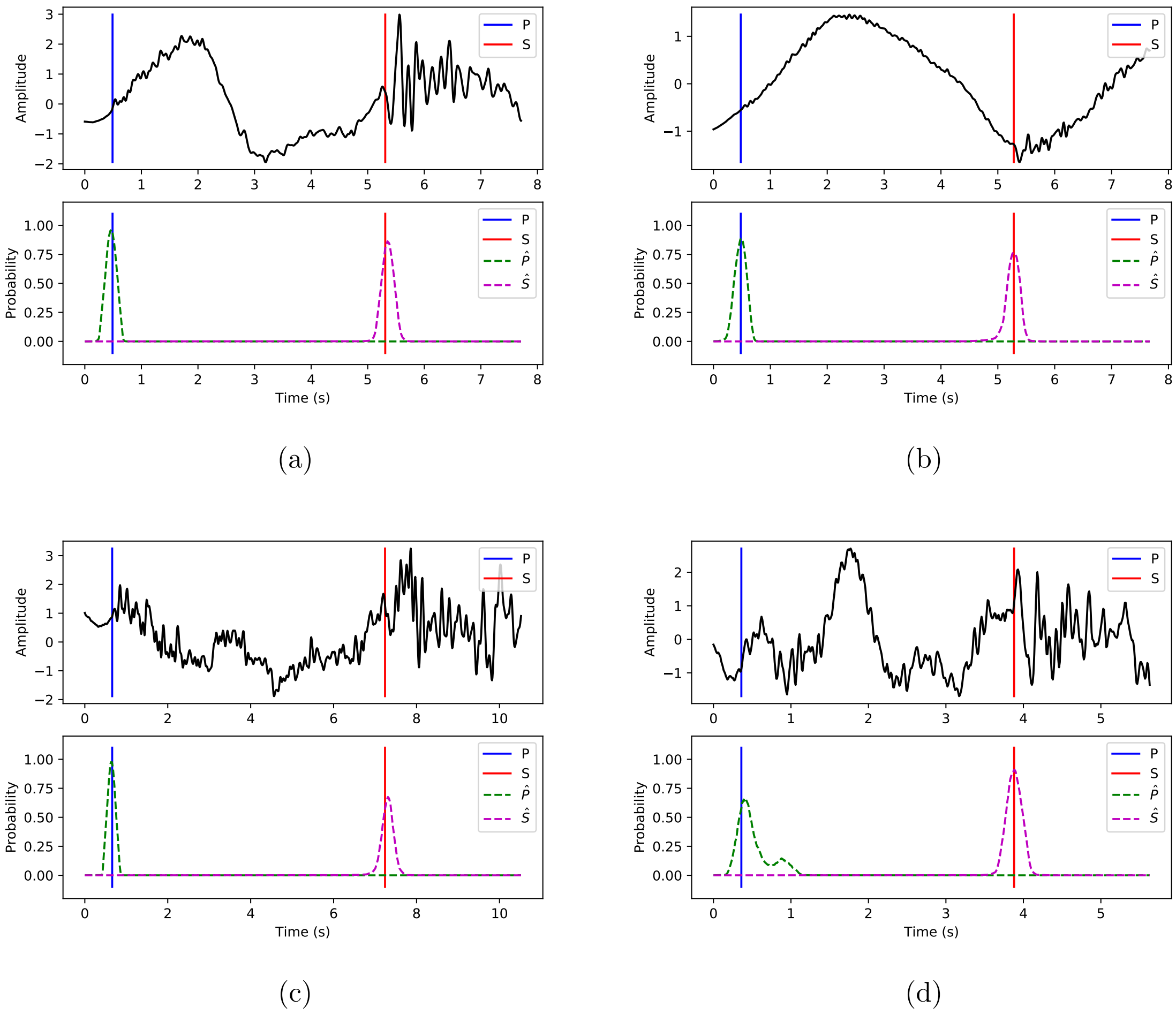

**Figure 15.** Examples with background variation

PhaseNet is not constrained by the input length or the number of earthquakes in a time window (Figure 12). The convolution of PhaseNet is done by a short filter scanning through the input time series. We can apply PhaseNet to data of any length to generate a running probability distribution of P or S wave arrivals, which can be used as the basis of earthquake detector when paired with an association algorithm. Accurate phase arrival times can also be used to get absolute earthquake locations and to develop seismic velocity models. PhaseNet provides an improved method to get accurate S arrivals, which will be useful for developing better S-wave velocity models and improving earthquake locations.

## 6 CONCLUSION

Deep learning methods are improving rapidly. An important ingredient for improving them is the existence of large labeled data sets. In seismology, we are fortunate to have such large data sets ready at hand in the form of decades of arrival times with accompanying waveforms. We are on the verge of, or perhaps have already arrived at, a threshold where neural networks are "superhuman" in the sense that they can outperform human analysts. In this paper, we have built a training dataset using manually picked P and S arrival times from the Northern California Seismic Network catalog. We have developed PhaseNet, a deep neural network algorithm that uses three component waveform data to predict the probability distribution of P waves, S waves, and noise. We extract arrival times from the peaks of these distributions. Test results show that our method achieves significant improvements compared with existing methods, particularly for S waves. PCA visualization shows that the condensed neural weights contain characteristics that allow the separation of P waves, S waves, and noise. While further testing against existing methods is required, we are not far from making such a capability operational. An increase in accurate P and S arrival-times will help us to continue to extract as much information as possible from rapidly growing waveform data sets for earthquake monitoring, and the ability to extract reliable S waves will allow us to improve shear wave velocity models substantially, which will be especially useful for prediction of path effects in strong ground motion prediction. Finally, we note that PhaseNet can also be used for other phases for which manually labeled training dataset are available.

## ACKNOWLEDGEMENTS

We thank Lind S. Gee and Stephane Zuzlewski for their help on downloading and processing the catalog and waveform data from NCEDC. We thank Seyed Mostafa Mousavi, Yixiao Sheng and Clara Yoon for helpful discussions. Waveform data, metadata, or data products for this study were accessed through the Northern California Earthquake Data Center (NCEDC). This research is supported by National Science Foundation (NSF) grant number EAR-1551462.